\documentclass[prl,twocolumn,amsmath,amssymb,amsfonts,showkeys,showpacs]{revtex4}
\usepackage{bm,graphicx,xcolor,subfigure,hyperref,multirow,rotating}
\newcommand{\mc}{\multicolumn}
\newcommand{\mr}{\multirow}
\newcommand{\md}{\mathcal{D}}
\begin{document}
\bibliographystyle{apsrev}

	\title{Two electrons on a hypersphere: a quasi-exactly solvable model}
	
	\author{Pierre-Fran\c{c}ois Loos}
	\affiliation{Research School of Chemistry, Australian National University, Canberra, Australian Capital Territory 0200, Australia}

	\author{Peter M. W. Gill}
 	\thanks{Corresponding author}
	\email{peter.gill@anu.edu.au}
	\affiliation{Research School of Chemistry, Australian National University, Canberra, Australian Capital Territory 0200, Australia}

	\date{\today}

\begin{abstract}
We show that the exact wave function for two electrons, interacting through a Coulomb potential but constrained to remain on the surface of a $\md$-sphere ($\md \ge 1$), is a polynomial in the interelectronic distance $u$ for a countably infinite set of values of the radius $R$.  A selection of these radii, and the associated energies, are reported for ground and excited states on the singlet and triplet manifolds.  We conclude that the $\md=3$ model bears the greatest similarity to normal physical systems.
\end{abstract}


	\maketitle

Quantum mechanical models for which it is possible to solve explicitly for a finite portion of the energy spectrum are said to be quasi-exactly solvable \cite{Ushveridze}.  They have ongoing value and are useful both for illuminating more complicated systems and for testing and developing theoretical approaches, such as density functional theory (DFT) \cite{HohenbergPRB1964, KohnPRA1965, ParrYang} and explicitly correlated methods \cite{KutzelniggTheorChemAcc1985, KutzelniggJChemPhys1991, HendersonPRA2004, BokhanPCCP2008}.  One of the most famous two-body models is the Hooke's law atom which consists of a pair of electrons, repelling Coulombically but trapped in a harmonic external potential with force constant $k$.  This system was first considered nearly 50 years ago by Kestner and Sinanoglu \cite{KestnerPhysRev1962}, solved analytically in 1989 for one particular $k$ value \cite{KaisJCP1989}, and later for a countably infinite set of $k$ values \cite{TautPRA1993}.

A related system consists of two electrons trapped on the surface of a sphere of radius $R$.  This has been used by Berry and collaborators \cite{EzraPRA1982, EzraPRA1983, OjhaPRA1987, HindePRA1990} to understand both weakly and strongly correlated systems and to suggest an ``alternating'' version of Hund's rule \cite{WarnerNature1985}.  Seidl utilized this system to develop new correlation functionals \cite{SeidlPRL2000} within the adiabatic connection in DFT \cite{SeidlPRA2007b}.  We will use the term ``spherium'' to describe this system.

In recent work \cite{LoosPRA2009}, we examined various schemes and described a method for obtaining near-exact estimates of the $^1S$ ground state energy of spherium for any given $R$.  Because the corresponding Hartree-Fock (HF) energies are also known exactly \cite{LoosPRA2009}, this is now one of the most complete theoretical models for understanding electron correlation effects.

In this Letter, we consider $\md$-spherium, the generalization in which the two electrons are trapped on a $\md$-sphere of radius $R$.  We adopt the convention that a $\md$-sphere is the surface of a ($\md+1$)-dimensional ball.  (Thus, for example, the Berry system is 2-spherium.)  We show that the Schr\"odinger equation for the $^1S$ and the $^3P$ states can be solved exactly for a countably infinite set of $R$ values and that the resulting wave functions are polynomials in the interelectronic distance $u = |\bm{r}_1-\bm{r}_2|$.  Other spin and angular momentum states can be addressed in the same way using the ansatz derived by Breit \cite{BreitPR1930}.

The electronic Hamiltonian, in atomic units, is
\begin{equation}
	\hat{H} = - \frac{\nabla_1^2}{2} - \frac{\nabla_2^2}{2} + \frac{1}{u}
\end{equation}
and because each electron moves on a $\md$-sphere, it is natural to adopt hyperspherical coordinates \cite{Louck60, KnirkPRL1974}.

For $^1S$ states, it can be then shown \cite{LoosPRA2009} that the wave function $S(u)$ satisfies the Schr{\"o}dinger equation
\begin{equation} \label{S-singlet}
	\left[ \frac{u^2}{4R^2} - 1 \right] \frac{d^2S}{du^2} + \left[ \frac{(2\md-1)u}{4R^2} - \frac{\md-1}{u} \right] \frac{dS}{du} + \frac{S}{u} = E S
\end{equation}
By introducing the dimensionless variable $x = u/2R$, this becomes a Heun equation \cite{Ronveaux} with singular points at $x = -1, 0, +1$.  Based on our previous work \cite{LoosPRA2009} and the known solutions of the Heun equation \cite{Polyanin}, we seek wave functions of the form
\begin{equation} \label{S_series}
	S(u) = \sum_{k=0}^\infty s_k\,u^k
\end{equation}
and substitution into \eqref{S-singlet} yields the recurrence relation
\begin{equation} \label{recurrence-singlet}
	s_{k+2} = \frac{ s_{k+1} + \left[ k(k+2\md-2) \frac{1}{4R^2} - E \right] s_k }{(k+2)(k+\md)}
\end{equation}
with the starting values
\begin{equation}
	\{s_0,s_1\} =	\begin{cases}
						\{0,1\}			&	\md  =   1	\\
						\{1,1/(\md-1)\}	&	\md \ge 2
					\end{cases}
\end{equation}
Thus, the Kato cusp conditions \cite{Kato1957} are
\begin{align}\label{cusp-circle}
	S(0) & = 0	&	\frac{S''(0)}{S'(0)} & = 1
\end{align}
for electrons on a circle ($\md=1$) and
\begin{equation} \label{S-cusp}
	\frac{S'(0)}{S(0)} = \frac{1}{\md-1}
\end{equation}
in higher dimensions.  We note that the ``normal'' Kato value of 1/2 arises for $\md=3$, suggesting that this may the most appropriate model for atomic or molecular systems.  We will return to this point below. 

The wave function \eqref{S_series} reduces to the polynomial
\begin{equation}
	S_{n,m}(u) = \sum_{k=0}^n s_k\,u^k
\end{equation}
(where $m$ the number of roots between $0$ and $2R$) if, and only if, $s_{n+1} = s_{n+2} = 0$.  Thus, the energy $E_{n,m}$ is a root of the polynomial equation $s_{n+1} = 0$ (where $\deg s_{n+1} = \lfloor (n+1)/2 \rfloor$) and the corresponding radius $R_{n,m}$ is found from \eqref{recurrence-singlet} which yields
\begin{equation} \label{E_S}
	R_{n,m}^2 E_{n,m} = \frac{n}{2}\left(\frac{n}{2}+\md-1\right)
\end{equation}
$S_{n,m}(u)$ is the exact wave function of the $m$-th excited state of $^1S$ symmetry for the radius $R_{n,m}$.

\begin{table}
\caption{\label{tab:lowest} Radius $R$, energy $E$ and wave function $S(u)$ or $T(u)$ of the first $^1S$ and $^3P$ polynomial solutions for two electrons on a $\md$-sphere}
\begin{ruledtabular}
\begin{tabular}{ccccc}
		State		&	$\md$	&	$2R$			&	$E$	&		$S(u)$ or $T(u)$	\\
\hline
\mr{4}{*}{$^1S$}	&	1		&	$\sqrt{6}$		&	2/3		&		$u(1 + u/2)$		\\
					&	2		&	$\sqrt{3}$		&	1		&		$1 + u$				\\
					&	3		&	$\sqrt{10}$		&	1/2		&		$1 + u/2$			\\
					&	4		&	$\sqrt{21}$		&	1/3		&		$1 + u/3$			\\
\hline
\mr{4}{*}{$^3P$}	&	1		&	$\sqrt{6}$		&	1/2		&		$1 + u/2$			\\
					&	2		&	$\sqrt{15}$ 		&	1/3		&		$1 + u/3$			\\
					&	3		&	$\sqrt{28}$		&	1/4		&		$1 + u/4$			\\
					&	4		&	$\sqrt{45}$		&	1/5		&		$1 + u/5$			\\
\end{tabular}
\end{ruledtabular}
\end{table}

If we write the $^3P$ state wave function as \cite{BreitPR1930}
\begin{equation}
	^3\Psi = (\cos \theta_1 - \cos \theta_2)\,T(u)
\end{equation}
where $\theta_1$ and $\theta_2$ are the $\md$-th hyperspherical angles of the two electrons \cite{Louck60, KnirkPRL1974}, the symmetric part satisfies the Schr{\"o}dinger equation
\begin{equation} \label{S-triplet}
	\left[ \frac{u^2}{4R^2} - 1 \right] \frac{d^2T}{du^2} + \left[ \frac{(2\md+1)u}{4R^2} - \frac{\md+1}{u} \right] \frac{dT}{du} + \frac{T}{u} = E T
\end{equation}
and the antisymmetric part provides an additional kinetic energy contribution $\md/(2R^2)$.

Substituting the power series expansion
\begin{equation} \label{T_series}
	T(u) = \sum_{k=0}^\infty t_k\,u^k
\end{equation}
into \eqref{S-triplet} yields the recurrence relation
\begin{equation} \label{recurrence-triplet}
	t_{k+2} = \frac{ t_{k+1} + \left[ k(k+2\md) \frac{1}{4R^2} - E \right] t_k }{(k+2)(k+\md+2)}
\end{equation}
with the starting values
\begin{equation}
	\{t_0,t_1\} = \{1, 1/(\md+1)\}
\end{equation}
yielding the cusp condition
\begin{equation} \label{T-cusp}
	\frac{T'(0)}{T(0)} = \frac{1}{\md+1}
\end{equation}

The wave function \eqref{T_series} reduces to the polynomial
\begin{equation}
	T_{n,m}(u) = \sum_{k=0}^n t_k\,u^k
\end{equation}
when the energy $E_{n,m}$ is a root of $t_{n+1} = 0$ and the corresponding radius $R_{n,m}$ is found from \eqref{recurrence-triplet} which yields
\begin{equation} \label{E_T}
	R_{n,m}^2 E_{n,m} = \frac{n}{2}\left(\frac{n}{2}+\md\right)
\end{equation}
$T_{n,m}(u)$ is the exact wave function of the $m$-th excited state of $^3P$ symmetry for the radius $R_{n,m}$.

It is illuminating to begin by examining the simplest $^1S$ and $^3P$ polynomial solutions.  Except in the $\md=1$ case, the first $^1S$ solution has
\begin{align}
	R_{1,0} & = \sqrt{\frac{(2\md-1)(2\md-2)}{8}}	&	E_{1,0} & = \frac{1}{\md-1}
\end{align}
and the first $^3P$ solution has
\begin{align}
	R_{1,0} & = \sqrt{\frac{(2\md+1)(2\md+2)}{8}}	&	E_{1,0} & = \frac{1}{\md+1}
\end{align}
These are tabulated for $\md = 1, 2, 3, 4$, together with the associated wave functions, in Table \ref{tab:lowest}.

\begin{table*}
\caption{\label{tab:TEOAS-singlet} Radii $R_{n,m}$ and energies $E_{n,m}$ for $^1S$ states of two electrons on a $\md$-sphere ($\md$=1,2,3)}
\begin{ruledtabular}
\begin{tabular}{cccccccccccccc}
&				&	\mc{4}{c}{$\md = 1$}		&	\mc{4}{c}{$\md = 2$}		&	\mc{4}{c}{$\md = 3$}		\\
						\cline{3-6}						\cline{7-10}						\cline{11-14}
&	$n$/$m$	&	0	&	1	&	2	&	3	&	0	&	1	&	2	&	3	&	0	&	1	&	2	&	3	\\
\hline
\mr{9}{*}{\rotatebox{90}{Radius}}                                                                                                                                                                                              
&	1	&			&			&			&			&	0.8660	&			&			&
														&	1.5811	&			&			&				\\
&	2	&	1.2247	&			&			&			&	2.6458	&			&			&
														&	4.0620	&			&			&				\\
&	3	&	3.3912	&			&			&			&	5.4312	&	1.4150	&			&
														&	7.5154	&	2.2404	&			&				\\
&	4	&	6.5439	&	1.9178	&			&			&	9.2211	&	3.7379	&			&
														&	11.961	&	5.3320	&			&				\\
&	5	&	10.693	&	4.7071	&			&			&	14.012	&	7.0848	&	1.9256	&
														&	17.404	&	9.3775	&	2.8554	&				\\
&	6	&	15.841	&	8.4583	&	2.5522	&			&	19.804	&	11.448	&	4.7683	&
														&	23.846	&	14.410	&	6.5350	&				\\
&	7	&	21.989	&	13.199	&	5.9404	&			&	26.597	&	16.817	&	8.6593	&	2.4123
														&	31.287	&	20.439	&	11.158	&	3.4415		\\
&	8	&	29.136	&	18.936	&	10.277	&	3.1515	&	34.389	&	23.190	&	13.583	&	5.7566
														&	39.728	&	27.466	&	16.768	&	7.6903		\\
\hline
\mr{9}{*}{\rotatebox{90}{Energy}}
&	1	&			&			&			&			&	1.0000	&			&			&
														&	0.5000	&			&			&				\\
&	2	&	0.6667 	&			&			&			&	0.2857	&			&			&
														&	0.1818	&			&			&				\\	
&	3	&	0.1957 	&			&			&			&	0.1271	&	1.8729	&			&
														&	0.0930	&	1.0459	&			&				\\
&	4	&	0.0934	&	1.0875	&			&			&	0.0706	&	0.4294	&			&
														&	0.0559	&	0.2814	&			&				\\
&	5	&	0.0547	&	0.2821	&			&			&	0.0446	&	0.1743	&	2.3597	&
														&	0.0371	&	0.1279	&	1.3798	&				\\
&	6	&	0.0359	&	0.1258	&	1.3817	&			&	0.0306	&	0.0916	&	0.5278	&
														&	0.0264	&	0.0722	&	0.3512	&				\\
&	7	&	0.0253	&	0.0703	&	0.3471	&			&	0.0223	&	0.0557	&	0.2100	&	2.7065
														&	0.0197	&	0.0461	&	0.1546	&	1.6253		\\
&	8	&	0.0188	&	0.0446	&	0.1515	&	1.6110	&	0.0169	&	0.0372	&	0.1084	&	0.6035
														&	0.0152	&	0.0318	&	0.0854	&	0.4058		\\
\end{tabular}
\end{ruledtabular}
\end{table*}

\begin{table*}
\caption{\label{tab:TEOAS-triplet} Radii $R_{n,m}$ and energies $E_{n,m}$ for $^3P$ states of two electrons on a $\md$-sphere ($\md$=1,2,3)}
\begin{ruledtabular}
\begin{tabular}{cccccccccccccc}
&				&	\mc{4}{c}{$\md = 1$}		&	\mc{4}{c}{$\md = 2$}		&	\mc{4}{c}{$\md = 3$}		\\
						\cline{3-6}						\cline{7-10}						\cline{11-14}
&	$n$/$m$	&	0	&	1	&	2	&	3	&	0	&	1	&	2	&	3	&	0	&	1	&	2	&	3	\\
\hline
\mr{9}{*}{\rotatebox{90}{Radius}}
&		1		&	1.2247	&			&			&			&	1.9365	&			&			&
																&	2.6458	&			&			&			\\
&		2		&	3.3912	&			&			&			&	4.7958	&			&			&
																&	6.2048	&			&			&			\\
&		3		&	6.5439	&	1.9178	&			&			&	8.6227	&	2.6738	&			&
																&	10.718	&	3.4111	&			&			\\
&		4		&	10.693	&	4.7071	&			&			&	13.435	&	6.2041	&			&
																&	16.205	&	7.6748	&			&			\\
&		5		&	15.841	&	8.4583	&	2.5522	&			&	19.241	&	10.665	&	3.3588	&
																&	22.678	&	12.852	&	4.1285	&			\\
&		6		&	21.989	&	13.199	&	5.9404	&			&	26.043	&	16.094	&	7.5340	&
																&	30.142	&	18.979	&	9.0701	&			\\
&		7		&	29.136	&	18.936	&	10.277	&	3.1515	&	33.842	&	22.505	&	12.615	&	4.0095
																&	38.600	&	26.077	&	14.897	&	4.8130	\\
&		8		&	37.283	&	25.671	&	15.599	&	7.1177	&	42.640	&	29.907	&	18.650	&	8.8083
																&	48.054	&	34.155	&	21.654	&	10.411	\\      
\hline
\mr{9}{*}{\rotatebox{90}{Energy}}
&		1		&	0.5000	&			&			&			&	0.3333	&			&			&
																&	0.2500	&			&			&			\\
&		2		&	0.1739	&			&			&			&	0.1304	&			&			&
																&	0.1039	&			&			&			\\
&		3		&	0.0876	&	1.0196	&			&			&	0.0706	&	0.7343	&			&
																&	0.0588	&	0.5801	&			&			\\
&		4		&	0.0525	&	0.2708	&			&			&	0.0443	&	0.2078	&			&
																&	0.0381	&	0.1698	&			&			\\
&		5		&	0.0349	&	0.1223	&	1.3433	&			&	0.0304	&	0.0989	&	0.9972	&
																&	0.0267	&	0.0832	&	0.8067	&			\\
&		6		&	0.0248	&	0.0689	&	0.3401	&			&	0.0221	&	0.0579	&	0.2643	&
																&	0.0198	&	0.0500	&	0.2188	&			\\
&		7		&	0.0186	&	0.0439	&	0.1491	&	1.5858	&	0.0168	&	0.0380	&	0.1210	&	1.1974
																&	0.0153	&	0.0335	&	0.1025	&	0.9821	\\
&		8		&	0.0144	&	0.0303	&	0.0822	&	0.3948	&	0.0132	&	0.0268	&	0.0690	&	0.3093
																&	0.0121	&	0.0240	&	0.0597	&	0.2583	\\      
\end{tabular}
\end{ruledtabular}
\end{table*}

In the $\md=1$ case (\textit{i.e.}~two electrons on a circle), the first singlet and triplet solutions have $E_{2,0} = 2/3$ and $E_{1,0} = 1/2$, respectively, for the same value of the radius ($\sqrt{6}/2 \approx 1.2247$). The corresponding wave functions are related by $S_{2,0} = u\,T_{1,0}$.  Unlike $T_{1,0}$, the singlet wavefunction $S_{2,0}$ vanishes at $u = 0$, and exhibits a second-order cusp condition, as shown in \eqref{cusp-circle}.

For the 2-spherium ($\md=2$ case), we know from our previous work \cite{LoosPRA2009} that the HF energy of the lowest $^1S$ state is $E_{\rm HF} = 1/R$.  It follows that the exact correlation energy for $R = \sqrt{3}/2$ is $E_{\rm corr} = 1-2/\sqrt{3} \approx -0.1547$ which is much larger than the limiting correlation energies of the helium-like ions ($-0.0467$) \cite{BakerPRA1990} or Hooke's law atoms ($-0.0497$) \cite{GillJCP2005}.  This confirms our view that electron correlation on the surface of a sphere is qualitatively different from that in three-dimensional physical space.

The 3-spherium ($\md=3$ case), in contrast, possesses the same singlet and triplet cusp conditions --- Eqs. \eqref{S-cusp} and \eqref{T-cusp} --- as those for electrons moving in three-dimensional physical space.  Indeed, the wave functions in Table \ref{tab:lowest}
\begin{align}
	S_{1,0}(u) & = 1 + u/2	&	(R & = \sqrt{5/2})	\\
	T_{1,0}(u) & = 1 + u/4	&	(R & = \sqrt{7})
\end{align}
have precisely the form of the ansatz used in Kutzelnigg's increasingly popular R12 methods \cite{KutzelniggTheorChemAcc1985, KutzelniggJChemPhys1991}.  Moreover, it can be shown \cite{SmallR} that, as $R \to 0$, the correlation energy $E_{\rm corr}$ approaches $-0.0476$, which nestles between the corresponding values for the helium-like ions ($-0.0467$) \cite{BakerPRA1990} and the Hooke's law atom ($-0.0497$) \cite{GillJCP2005}.  Again, this suggests that the $\md=3$ model (``electrons on a hypersphere'') bears more similarity to common physical systems than the $\md=2$ model (``electrons on a sphere'').

Numerical values of the energies and radii, for polynomial wave functions in $\md=1, 2, 3$, are reported in Table \ref{tab:TEOAS-singlet} (for $^1S$ states) and Table \ref{tab:TEOAS-triplet} (for $^3P$ states).

For fixed $\md$, the radii increase with $n$ but decrease with $m$, and the energies behave in exactly the opposite way.  As $R$ (or equivalently $n$) increases, the electrons tend to localize on opposite sides of the sphere, a phenomenon known as Wigner crystallization \cite{WignerPR1934} which has also been observed in other systems \cite{ThompsonPRB2004,TautPRA1993}.  As a result, for large $R$, the ground state energies of both the singlet and triplet state approach $1/(2R)$.  Analogous behavior is observed when $\md \to \infty$ \cite{YaffeRMP1982,GoodsonJCP1987}.

In conclusion, we have shown that the system of two electrons, interacting via a Coulomb potential but constrained to remain on a $\md$-sphere, can be solved exactly for an infinite set of values of the radius $R$.  We find that the 3-spherium ($\md=3$ model), wherein the electrons are confined to a three-dimensional surface of a four-dimensional ball, has greater similarity to normal physical systems than the more familiar $\md=2$ case.

We believe that our results will be useful in the future development of correlation functionals within density-functional theory \cite{GoriGiorgiJCTC2009}, intracule functional theory \cite{GillPCCP2006, DumontPCCP2007, CrittendenJCP2007a, CrittendenJCP2007b, BernardPCCP2008, PearsonJCP2009}, and explicitly correlated methods \cite{KutzelniggTheorChemAcc1985, KutzelniggJChemPhys1991, HendersonPRA2004, BokhanPCCP2008}.  They also shed new light on dimension-dependent correlation effects, and may be used as an alternative system for studying quantum dots \cite{HendersonCPL2001}.

PMWG thanks the APAC Merit Allocation Scheme for a grant of supercomputer time and the Australian Research Council (Grant DP0664466) for funding.


\begin{thebibliography}{40}
\expandafter\ifx\csname natexlab\endcsname\relax\def\natexlab#1{#1}\fi
\expandafter\ifx\csname bibnamefont\endcsname\relax
  \def\bibnamefont#1{#1}\fi
\expandafter\ifx\csname bibfnamefont\endcsname\relax
  \def\bibfnamefont#1{#1}\fi
\expandafter\ifx\csname citenamefont\endcsname\relax
  \def\citenamefont#1{#1}\fi
\expandafter\ifx\csname url\endcsname\relax
  \def\url#1{\texttt{#1}}\fi
\expandafter\ifx\csname urlprefix\endcsname\relax\def\urlprefix{URL }\fi
\providecommand{\bibinfo}[2]{#2}
\providecommand{\eprint}[2][]{\url{#2}}

\bibitem[{\citenamefont{Ushveridze}(1994)}]{Ushveridze}
\bibinfo{author}{\bibfnamefont{A.~G.} \bibnamefont{Ushveridze}},
  \emph{\bibinfo{title}{Quasi-Exactly Solvable Models in Quantum Mechanics}}
  (\bibinfo{publisher}{Institute of Physics Publishing}, \bibinfo{year}{1994}).

\bibitem[{\citenamefont{Hohenberg and Kohn}(1964)}]{HohenbergPRB1964}
\bibinfo{author}{\bibfnamefont{P.}~\bibnamefont{Hohenberg}} \bibnamefont{and}
  \bibinfo{author}{\bibfnamefont{W.}~\bibnamefont{Kohn}},
  \bibinfo{journal}{Phys. Rev.} \textbf{\bibinfo{volume}{136}},
  \bibinfo{pages}{B864} (\bibinfo{year}{1964}).

\bibitem[{\citenamefont{Kohn and Sham}(1965)}]{KohnPRA1965}
\bibinfo{author}{\bibfnamefont{W.}~\bibnamefont{Kohn}} \bibnamefont{and}
  \bibinfo{author}{\bibfnamefont{L.}~\bibnamefont{Sham}},
  \bibinfo{journal}{Phys. Rev.} \textbf{\bibinfo{volume}{140}},
  \bibinfo{pages}{A1133} (\bibinfo{year}{1965}).

\bibitem[{\citenamefont{Parr and Yang}(1989)}]{ParrYang}
\bibinfo{author}{\bibfnamefont{R.~G.} \bibnamefont{Parr}} \bibnamefont{and}
  \bibinfo{author}{\bibfnamefont{W.}~\bibnamefont{Yang}},
  \emph{\bibinfo{title}{Density Functional Theory for Atoms and Molecules}}
  (\bibinfo{publisher}{Oxford University Press}, \bibinfo{year}{1989}).

\bibitem[{\citenamefont{Kutzelnigg}(1985)}]{KutzelniggTheorChemAcc1985}
\bibinfo{author}{\bibfnamefont{W.}~\bibnamefont{Kutzelnigg}},
  \bibinfo{journal}{Theor. Chim. Acta} \textbf{\bibinfo{volume}{68}},
  \bibinfo{pages}{445} (\bibinfo{year}{1985}).

\bibitem[{\citenamefont{Kutzelnigg and
  Klopper}(1991)}]{KutzelniggJChemPhys1991}
\bibinfo{author}{\bibfnamefont{W.}~\bibnamefont{Kutzelnigg}} \bibnamefont{and}
  \bibinfo{author}{\bibfnamefont{W.}~\bibnamefont{Klopper}},
  \bibinfo{journal}{J. Chem. Phys.} \textbf{\bibinfo{volume}{94}},
  \bibinfo{pages}{1985} (\bibinfo{year}{1991}).

\bibitem[{\citenamefont{Henderson and Bartlett}(2004)}]{HendersonPRA2004}
\bibinfo{author}{\bibfnamefont{T.~M.} \bibnamefont{Henderson}}
  \bibnamefont{and} \bibinfo{author}{\bibfnamefont{R.~J.}
  \bibnamefont{Bartlett}}, \bibinfo{journal}{Phys. Rev. A}
  \textbf{\bibinfo{volume}{70}}, \bibinfo{pages}{022512}
  (\bibinfo{year}{2004}).

\bibitem[{\citenamefont{Bokhan et~al.}(2008)\citenamefont{Bokhan, Ten-no, and
  Noga}}]{BokhanPCCP2008}
\bibinfo{author}{\bibfnamefont{D.}~\bibnamefont{Bokhan}},
  \bibinfo{author}{\bibfnamefont{S.}~\bibnamefont{Ten-no}}, \bibnamefont{and}
  \bibinfo{author}{\bibfnamefont{J.}~\bibnamefont{Noga}},
  \bibinfo{journal}{Phys. Chem. Chem. Phys.} \textbf{\bibinfo{volume}{10}},
  \bibinfo{pages}{3320} (\bibinfo{year}{2008}).

\bibitem[{\citenamefont{Kestner and Sinanoglu}(1962)}]{KestnerPhysRev1962}
\bibinfo{author}{\bibfnamefont{N.~R.} \bibnamefont{Kestner}} \bibnamefont{and}
  \bibinfo{author}{\bibfnamefont{O.}~\bibnamefont{Sinanoglu}},
  \bibinfo{journal}{Phys. Rev.} \textbf{\bibinfo{volume}{128}},
  \bibinfo{pages}{2687} (\bibinfo{year}{1962}).

\bibitem[{\citenamefont{Kais et~al.}(1989)\citenamefont{Kais, Herschbach, and
  Levine}}]{KaisJCP1989}
\bibinfo{author}{\bibfnamefont{S.}~\bibnamefont{Kais}},
  \bibinfo{author}{\bibfnamefont{D.~R.} \bibnamefont{Herschbach}},
  \bibnamefont{and} \bibinfo{author}{\bibfnamefont{R.~D.}
  \bibnamefont{Levine}}, \bibinfo{journal}{J. Chem. Phys}
  \textbf{\bibinfo{volume}{91}}, \bibinfo{pages}{7791} (\bibinfo{year}{1989}).

\bibitem[{\citenamefont{Taut}(1993)}]{TautPRA1993}
\bibinfo{author}{\bibfnamefont{M.}~\bibnamefont{Taut}}, \bibinfo{journal}{Phys.
  Rev. A} \textbf{\bibinfo{volume}{48}}, \bibinfo{pages}{3561}
  (\bibinfo{year}{1993}).

\bibitem[{\citenamefont{Ezra and Berry}(1982)}]{EzraPRA1982}
\bibinfo{author}{\bibfnamefont{G.~S.} \bibnamefont{Ezra}} \bibnamefont{and}
  \bibinfo{author}{\bibfnamefont{R.~S.} \bibnamefont{Berry}},
  \bibinfo{journal}{Phys. Rev. A} \textbf{\bibinfo{volume}{25}},
  \bibinfo{pages}{1513} (\bibinfo{year}{1982}).

\bibitem[{\citenamefont{Ezra and Berry}(1983)}]{EzraPRA1983}
\bibinfo{author}{\bibfnamefont{G.~S.} \bibnamefont{Ezra}} \bibnamefont{and}
  \bibinfo{author}{\bibfnamefont{R.~S.} \bibnamefont{Berry}},
  \bibinfo{journal}{Phys. Rev. A} \textbf{\bibinfo{volume}{28}},
  \bibinfo{pages}{1989} (\bibinfo{year}{1983}).

\bibitem[{\citenamefont{Ojha and Berry}(1987)}]{OjhaPRA1987}
\bibinfo{author}{\bibfnamefont{P.~C.} \bibnamefont{Ojha}} \bibnamefont{and}
  \bibinfo{author}{\bibfnamefont{R.~S.} \bibnamefont{Berry}},
  \bibinfo{journal}{Phys. Rev. A} \textbf{\bibinfo{volume}{36}},
  \bibinfo{pages}{1575} (\bibinfo{year}{1987}).

\bibitem[{\citenamefont{Hinde and Berry}(1990)}]{HindePRA1990}
\bibinfo{author}{\bibfnamefont{R.~J.} \bibnamefont{Hinde}} \bibnamefont{and}
  \bibinfo{author}{\bibfnamefont{R.~S.} \bibnamefont{Berry}},
  \bibinfo{journal}{Phys. Rev. A} \textbf{\bibinfo{volume}{42}},
  \bibinfo{pages}{2259} (\bibinfo{year}{1990}).

\bibitem[{\citenamefont{Warner and Berry}(1985)}]{WarnerNature1985}
\bibinfo{author}{\bibfnamefont{J.~W.} \bibnamefont{Warner}} \bibnamefont{and}
  \bibinfo{author}{\bibfnamefont{R.~S.} \bibnamefont{Berry}},
  \bibinfo{journal}{Nature} \textbf{\bibinfo{volume}{313}},
  \bibinfo{pages}{160} (\bibinfo{year}{1985}).

\bibitem[{\citenamefont{Seidl et~al.}(2000)\citenamefont{Seidl, Perdew, and
  Kurth}}]{SeidlPRL2000}
\bibinfo{author}{\bibfnamefont{M.}~\bibnamefont{Seidl}},
  \bibinfo{author}{\bibfnamefont{J.~P.} \bibnamefont{Perdew}},
  \bibnamefont{and} \bibinfo{author}{\bibfnamefont{S.}~\bibnamefont{Kurth}},
  \bibinfo{journal}{Phys. Rev. Lett.} \textbf{\bibinfo{volume}{84}},
  \bibinfo{pages}{5070} (\bibinfo{year}{2000}).

\bibitem[{\citenamefont{Seidl}(2007)}]{SeidlPRA2007b}
\bibinfo{author}{\bibfnamefont{M.}~\bibnamefont{Seidl}},
  \bibinfo{journal}{Phys. Rev. A} \textbf{\bibinfo{volume}{75}},
  \bibinfo{pages}{062506} (\bibinfo{year}{2007}).

\bibitem[{\citenamefont{Loos and Gill}(2009)}]{LoosPRA2009}
\bibinfo{author}{\bibfnamefont{P.-F.} \bibnamefont{Loos}} \bibnamefont{and}
  \bibinfo{author}{\bibfnamefont{P.~M.~W.} \bibnamefont{Gill}},
  \bibinfo{journal}{Phys. Rev. A} \textbf{\bibinfo{volume}{79}},
  \bibinfo{pages}{062517} (\bibinfo{year}{2009}).

\bibitem[{\citenamefont{Breit}(1930)}]{BreitPR1930}
\bibinfo{author}{\bibfnamefont{G.}~\bibnamefont{Breit}},
  \bibinfo{journal}{Phys. Rev.} \textbf{\bibinfo{volume}{35}},
  \bibinfo{pages}{569} (\bibinfo{year}{1930}).

\bibitem[{\citenamefont{Louck}(1960)}]{Louck60}
\bibinfo{author}{\bibfnamefont{J.~D.} \bibnamefont{Louck}},
  \bibinfo{journal}{J. Mol. Spectrosc.} \textbf{\bibinfo{volume}{4}},
  \bibinfo{pages}{298} (\bibinfo{year}{1960}).

\bibitem[{\citenamefont{Knirk}(1974)}]{KnirkPRL1974}
\bibinfo{author}{\bibfnamefont{D.~L.} \bibnamefont{Knirk}},
  \bibinfo{journal}{Phys. Rev. Lett.} \textbf{\bibinfo{volume}{32}},
  \bibinfo{pages}{651} (\bibinfo{year}{1974}).

\bibitem[{\citenamefont{Ronveaux}(1995)}]{Ronveaux}
\bibinfo{editor}{\bibfnamefont{A.}~\bibnamefont{Ronveaux}}, ed.,
  \emph{\bibinfo{title}{Heun's Differential Equations}}
  (\bibinfo{publisher}{Oxford University Press}, \bibinfo{address}{Oxford},
  \bibinfo{year}{1995}).

\bibitem[{\citenamefont{Polyanin and Zaitsev}(2003)}]{Polyanin}
\bibinfo{author}{\bibfnamefont{A.~D.} \bibnamefont{Polyanin}} \bibnamefont{and}
  \bibinfo{author}{\bibfnamefont{V.~F.} \bibnamefont{Zaitsev}},
  \emph{\bibinfo{title}{Handbook of Exact solutions for Differential
  Equations}} (\bibinfo{publisher}{Chapman \& Hall/CRC}, \bibinfo{year}{2003}).

\bibitem[{\citenamefont{Kato}(1957)}]{Kato1957}
\bibinfo{author}{\bibfnamefont{T.}~\bibnamefont{Kato}},
  \bibinfo{journal}{Commun. Pure Appl. Math.} \textbf{\bibinfo{volume}{10}},
  \bibinfo{pages}{151} (\bibinfo{year}{1957}).

\bibitem[{\citenamefont{Baker et~al.}(1990)\citenamefont{Baker, Freund, Hill,
  and {Morgan III}}}]{BakerPRA1990}
\bibinfo{author}{\bibfnamefont{J.~D.} \bibnamefont{Baker}},
  \bibinfo{author}{\bibfnamefont{D.~E.} \bibnamefont{Freund}},
  \bibinfo{author}{\bibfnamefont{R.~N.} \bibnamefont{Hill}}, \bibnamefont{and}
  \bibinfo{author}{\bibfnamefont{J.~D.} \bibnamefont{{Morgan III}}},
  \bibinfo{journal}{Phys. Rev. A} \textbf{\bibinfo{volume}{41}},
  \bibinfo{pages}{1241} (\bibinfo{year}{1990}).

\bibitem[{\citenamefont{Gill and O'Neill}(2005)}]{GillJCP2005}
\bibinfo{author}{\bibfnamefont{P.~M.~W.} \bibnamefont{Gill}} \bibnamefont{and}
  \bibinfo{author}{\bibfnamefont{D.~P.} \bibnamefont{O'Neill}},
  \bibinfo{journal}{J. Chem. Phys.} \textbf{\bibinfo{volume}{122}},
  \bibinfo{pages}{094110} (\bibinfo{year}{2005}).

\bibitem[{\citenamefont{Loos et~al.}(in preparation)\citenamefont{Loos,
  Gilbert, and Gill}}]{SmallR}
\bibinfo{author}{\bibfnamefont{P.~F.} \bibnamefont{Loos}},
  \bibinfo{author}{\bibfnamefont{A.~T.~B.} \bibnamefont{Gilbert}},
  \bibnamefont{and} \bibinfo{author}{\bibfnamefont{P.~M.~W.}
  \bibnamefont{Gill}} (\bibinfo{year}{in preparation}).

\bibitem[{\citenamefont{Wigner}(1934)}]{WignerPR1934}
\bibinfo{author}{\bibfnamefont{E.}~\bibnamefont{Wigner}},
  \bibinfo{journal}{Phys. Rev.} \textbf{\bibinfo{volume}{46}},
  \bibinfo{pages}{1002} (\bibinfo{year}{1934}).

\bibitem[{\citenamefont{Thompson and Alavi}(2004)}]{ThompsonPRB2004}
\bibinfo{author}{\bibfnamefont{D.~C.} \bibnamefont{Thompson}} \bibnamefont{and}
  \bibinfo{author}{\bibfnamefont{A.}~\bibnamefont{Alavi}},
  \bibinfo{journal}{Phys. Rev. B} \textbf{\bibinfo{volume}{69}},
  \bibinfo{pages}{201302} (\bibinfo{year}{2004}).

\bibitem[{\citenamefont{Yaffe}(1982)}]{YaffeRMP1982}
\bibinfo{author}{\bibfnamefont{L.~G.} \bibnamefont{Yaffe}},
  \bibinfo{journal}{Rev. Mod. Phys.} \textbf{\bibinfo{volume}{54}},
  \bibinfo{pages}{407} (\bibinfo{year}{1982}).

\bibitem[{\citenamefont{Goodson and Herschbach}(1987)}]{GoodsonJCP1987}
\bibinfo{author}{\bibfnamefont{D.~Z.} \bibnamefont{Goodson}} \bibnamefont{and}
  \bibinfo{author}{\bibfnamefont{D.~R.} \bibnamefont{Herschbach}},
  \bibinfo{journal}{J. Chem. Phys.} \textbf{\bibinfo{volume}{86}},
  \bibinfo{pages}{4997} (\bibinfo{year}{1987}).

\bibitem[{\citenamefont{Gori-Giorgi et~al.}(2009)\citenamefont{Gori-Giorgi,
  Vignale, and Seidl}}]{GoriGiorgiJCTC2009}
\bibinfo{author}{\bibfnamefont{P.}~\bibnamefont{Gori-Giorgi}},
  \bibinfo{author}{\bibfnamefont{G.}~\bibnamefont{Vignale}}, \bibnamefont{and}
  \bibinfo{author}{\bibfnamefont{M.}~\bibnamefont{Seidl}}, \bibinfo{journal}{J.
  Chem. Theor. Comput.} \textbf{\bibinfo{volume}{5}}, \bibinfo{pages}{743}
  (\bibinfo{year}{2009}).

\bibitem[{\citenamefont{Gill et~al.}(2006)\citenamefont{Gill, Crittenden,
  O'Neill, and Besley}}]{GillPCCP2006}
\bibinfo{author}{\bibfnamefont{P.~M.~W.} \bibnamefont{Gill}},
  \bibinfo{author}{\bibfnamefont{D.~L.} \bibnamefont{Crittenden}},
  \bibinfo{author}{\bibfnamefont{D.~P.} \bibnamefont{O'Neill}},
  \bibnamefont{and} \bibinfo{author}{\bibfnamefont{N.~A.}
  \bibnamefont{Besley}}, \bibinfo{journal}{Phys. Chem. Chem. Phys.}
  \textbf{\bibinfo{volume}{8}}, \bibinfo{pages}{15} (\bibinfo{year}{2006}).

\bibitem[{\citenamefont{Dumont et~al.}(2007)\citenamefont{Dumont, Crittenden,
  and Gill}}]{DumontPCCP2007}
\bibinfo{author}{\bibfnamefont{E.~E.} \bibnamefont{Dumont}},
  \bibinfo{author}{\bibfnamefont{D.~L.} \bibnamefont{Crittenden}},
  \bibnamefont{and} \bibinfo{author}{\bibfnamefont{P.~M.~W.}
  \bibnamefont{Gill}}, \bibinfo{journal}{Phys. Chem. Chem. Phys.}
  \textbf{\bibinfo{volume}{9}}, \bibinfo{pages}{5340} (\bibinfo{year}{2007}).

\bibitem[{\citenamefont{Crittenden and Gill}(2007)}]{CrittendenJCP2007a}
\bibinfo{author}{\bibfnamefont{D.~L.} \bibnamefont{Crittenden}}
  \bibnamefont{and} \bibinfo{author}{\bibfnamefont{P.~M.~W.}
  \bibnamefont{Gill}}, \bibinfo{journal}{J. Chem. Phys.}
  \textbf{\bibinfo{volume}{127}}, \bibinfo{pages}{014101}
  (\bibinfo{year}{2007}).

\bibitem[{\citenamefont{Crittenden et~al.}(2007)\citenamefont{Crittenden,
  Dumont, and Gill}}]{CrittendenJCP2007b}
\bibinfo{author}{\bibfnamefont{D.~L.} \bibnamefont{Crittenden}},
  \bibinfo{author}{\bibfnamefont{E.~E.} \bibnamefont{Dumont}},
  \bibnamefont{and} \bibinfo{author}{\bibfnamefont{P.~M.~W.}
  \bibnamefont{Gill}}, \bibinfo{journal}{J. Chem. Phys.}
  \textbf{\bibinfo{volume}{127}}, \bibinfo{pages}{141103}
  (\bibinfo{year}{2007}).

\bibitem[{\citenamefont{Bernard et~al.}(2008)\citenamefont{Bernard, Crittenden,
  and Gill}}]{BernardPCCP2008}
\bibinfo{author}{\bibfnamefont{Y.~A.} \bibnamefont{Bernard}},
  \bibinfo{author}{\bibfnamefont{D.~L.} \bibnamefont{Crittenden}},
  \bibnamefont{and} \bibinfo{author}{\bibfnamefont{P.~M.~W.}
  \bibnamefont{Gill}}, \bibinfo{journal}{Phys. Chem. Chem. Phys.}
  \textbf{\bibinfo{volume}{10}}, \bibinfo{pages}{3447} (\bibinfo{year}{2008}).

\bibitem[{\citenamefont{Pearson et~al.}(2009)\citenamefont{Pearson, Crittenden,
  and Gill}}]{PearsonJCP2009}
\bibinfo{author}{\bibfnamefont{J.~K.} \bibnamefont{Pearson}},
  \bibinfo{author}{\bibfnamefont{D.~L.} \bibnamefont{Crittenden}},
  \bibnamefont{and} \bibinfo{author}{\bibfnamefont{P.~M.~W.}
  \bibnamefont{Gill}}, \bibinfo{journal}{J. Chem. Phys.}
  \textbf{\bibinfo{volume}{130}}, \bibinfo{pages}{164110}
  (\bibinfo{year}{2009}).

\bibitem[{\citenamefont{Henderson et~al.}(2001)\citenamefont{Henderson, Runge,
  and Bartlett}}]{HendersonCPL2001}
\bibinfo{author}{\bibfnamefont{T.~M.} \bibnamefont{Henderson}},
  \bibinfo{author}{\bibfnamefont{K.}~\bibnamefont{Runge}}, \bibnamefont{and}
  \bibinfo{author}{\bibfnamefont{R.~J.} \bibnamefont{Bartlett}},
  \bibinfo{journal}{Chem. Phys. Lett.} \textbf{\bibinfo{volume}{337}},
  \bibinfo{pages}{138} (\bibinfo{year}{2001}).

\end{thebibliography}
\end{document}